\documentclass[aps,pra,amssymb,twocolumn,amsmath,superscriptaddress,showpacs,10pt]{revtex4-1}

\usepackage{graphicx}
\usepackage{dcolumn}
\usepackage{bm}
\usepackage{color}
\usepackage{epstopdf}
\usepackage{booktabs}

\def\e{\mathrm{e}}
\def\d{\mathrm{d}}
\def\i{\mathrm{i}}
\def\I{\mathcal{I}}

\begin{document}

\title{Exploring plenoptic properties of correlation imaging with chaotic light}

\author{Francesco V. Pepe}
\affiliation{Museo Storico della Fisica e Centro Studi e Ricerche ``Enrico Fermi'', I-00184 Roma, Italy} 
\affiliation{INFN, Sezione di Bari, I-70126 Bari, Italy}

\author{Ornella Vaccarelli} 
\affiliation{Dipartimento Interateneo di Fisica, Universit\`a degli studi di Bari, I-70126 Bari, Italy}
\affiliation{Institut de Min\'eralogie, de Physique des Mat\'eriaux et de Cosmochimie, Sorbonne Universit\'es, Universit\'e Pierre et Marie Curie - Paris 6, UMR CNRS 7590, Museum National d'Histoire Naturelle, IRD UMR 206, F-75005 Paris, France}

\author{Augusto Garuccio}
\affiliation{Dipartimento Interateneo di Fisica, Universit\`a degli studi di Bari, I-70126 Bari, Italy} 
\affiliation{INFN, Sezione di Bari, I-70126 Bari, Italy}
\affiliation{Istituto Nazionale di Ottica (INO-CNR), I-50125 Firenze, Italy}

\author{Giuliano Scarcelli}
\affiliation{Fischell Department of Bioengineering, University of Maryland, College Park MD 20742 USA} 

\author{Milena D'Angelo}\email{milena.dangelo@uniba.it}
\affiliation{Dipartimento Interateneo di Fisica, Universit\`a degli studi di Bari, I-70126 Bari, Italy} 
\affiliation{INFN, Sezione di Bari, I-70126 Bari, Italy}
\affiliation{Istituto Nazionale di Ottica (INO-CNR), I-50125 Firenze, Italy}

\begin{abstract}
In a setup illuminated by chaotic light, we consider different schemes that enable to perform imaging by measuring second-order intensity correlations. The most relevant feature of the proposed protocols is the ability to perform plenoptic imaging, namely to reconstruct the geometrical path of light propagating in the system, by imaging both the object and the focusing element. This property allows to encode, in a single data acquisition, both multi-perspective images of the scene and light distribution in different planes between the scene and the focusing element. We unveil the plenoptic property of three different setups, explore their refocusing potentialities and discuss their practical applications.
\end{abstract}


\maketitle

\section{Introduction}

Plenoptic imaging is a recently established optical method for recording, in a single exposure, both the planar distribution and the propagation direction of light. Though the intuition of plenoptic imaging goes back in 1908 \cite{lippmann}, the first feasible proposal of realization of a light field camera came in much more recent times \cite{adelson}. Plenoptic imaging was employed for the first time in enhanced digital cameras \cite{website,ng}, and currently has a wide range of scientific applications, that include stereoscopy \cite{adelson,muenzel,levoy}, microscopy \cite{microscopy1,microscopy2,microscopy3,microscopy4}, velocimetry \cite{piv}, tracking and sizing of particles \cite{tracking}, wavefront sensing \cite{thesis_wu,eye,atmosphere1,atmosphere2}. One of the most relevant features of plenoptic imaging is the simultaneous acquisition of multi-perspective 2D images, which enable a fast reconstruction of the 3D scene \cite{3dimaging}. Among the frontier applications, the plenoptic scheme has been employed in several uses, e.g.\ high-speed and large-scale 3D functional imaging of neuronal activity \cite{microscopy4}, eye aberration measurement and iris imaging \cite{eye,eye2}, first studies for surgical robotics \cite{surgery}, endoscopy \cite{endoscopy}, blood \cite{piv2} and air flow \cite{airflow1,airflow2} analysis. Recently, novel configurations, including plenoptic $2.0$ and multi-focused plenoptic, have been developed \cite{focused_pleno,multifocused_pleno,plenoptic_review, georgiev2009high,jin2017point}, as well as algorithms and analysis tools \cite{dansereau2013decoding,perez2014super,li2016scalable}.

\begin{figure}
\centering
\includegraphics[width=0.45\textwidth]{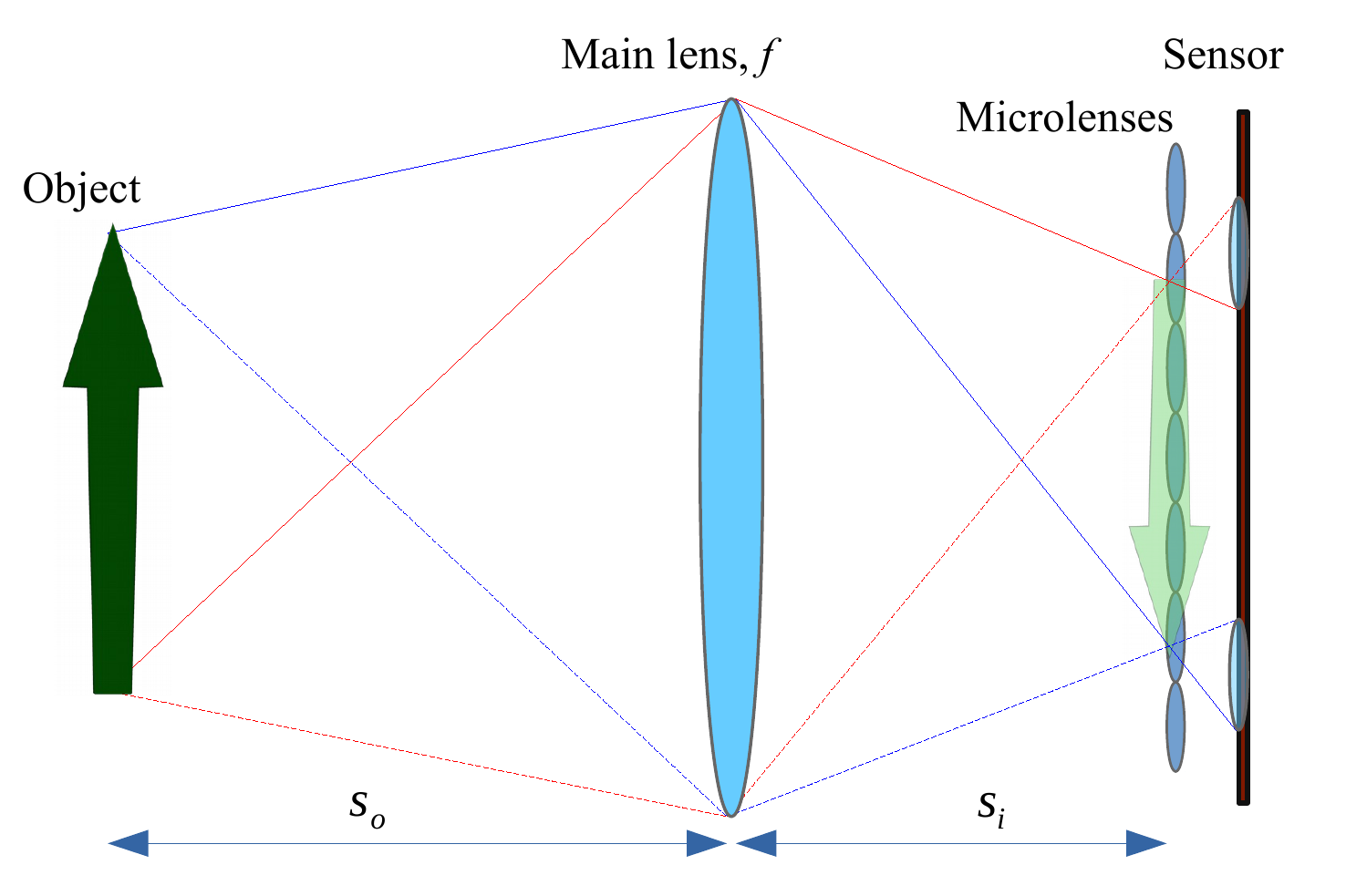}
\caption{Schematic representation of the structure of a standard plenoptic camera, whose peculiar coponent is the microlens array placed in front of the camera sensor. The main camera lens is such that, when the object-to-lens distance $s_o$ and the lens-to-microlenses distance $s_i$ satisfy the thin-lens equation ($1/s_i+1/s_o=1/f$, with $f$ the focal length of the main camera lens), the image of the object is focused on the microlens array. On the other hand, multiple images of the main lens are focused on the sensor by the microlenses;  the collection of such images enable to refocus planes different from the focused one.}\label{fig:plenoptic}
\end{figure}

A standard plenoptic camera, schematically represented in Figure \ref{fig:plenoptic}, is characterized by a microlens array placed in front of the sensor. Each microlens focuses an image of the main camera lens on a dedicated portion of the sensor, while the image of the scene forms on the microlens array. Hence, each microlens plays the role of an imaging pixel \cite{adelson, ng}. The multiple images of the main camera lens enable reconstructing the direction of light impinging on the camera; this information can be used to refocus different planes within the acquired image, extend its depth of field, and change the point of view on the scene. However, the natural tradeoff between resolution and depth of field is still present in plenoptic devices: collecting angular information by the use of microlenses limits the image resolution, that is defined by the microlens size rather than the pixel size. If $N_{\mathrm{tot}}$ is the total number of pixels per line on the sensor, $N_x$ the number of microlenses per line, and $N_u$ the number of pixels per line behind each microlens, then $N_x N_u = N_{\mathrm{tot}}$. Thus, $N_u$, that fixes the directional resolution, also determines the loss of image resolution, which is reduced to the one given by a lens having an $N_u$ times smaller numerical aperture. The final advantage of conventional plenoptic imaging is thus practical rather than fundamental, and is related to both the higher signal-to-noise ratio of the reconstructed image and to the parallel acquisition of images from many perspectives.

Recently, a new plenoptic scheme, based on correlated light sources and second-order correlation measurements, has been proposed for overcoming this fundamental limit; the scheme was named Correlation Plenoptic Imaging (CPI) and was demonstrated both for chaotic light \cite{cpi_prl,cpi_qmqm} and entangled photon pairs \cite{cpi_technologies}. Exploiting spatio-temporal correlation properties of light, spatial and directional detection can be performed on \textit{two distinct sensors}: one sensor captures the high-resolution ``ghost'' image of the object \cite{pittman,gatti,laserphys,valencia,scarcelliPRL}, while the other one detects an image of the focusing element, which enables to reconstruct the path of light in the setup, as in a standard plenoptic device. From a practical point of view, the relation between the spatial ($N_x$) and the angular ($N_u$) pixels per line, at fixed $N_{\mathrm{tot}}$, becomes linear rather than hyperbolic: $N_x + N_u = N_{tot}$ \cite{cpi_prl}. Such a novel approach to plenoptic imaging aims at keeping the advantages of the conventional approach without renouncing to diffraction-limited resolution, thus fostering promising practical applications.

In this paper, we will extend our analysis to unveil the plenoptic imaging properties of two different setups, having the common property to be able to retrieve a conventional first-order image of the object, rather than a ghost image. The available first-order image offers the possibility of choosing the acquisition method: a conventional camera, a standard plenoptic camera, CPI. This scenario turns out to be particularly useful whenever the two plenoptic techniques have competing performances in specific ranges of image resolution. The proposed schemes are also oriented to a direct experimental comparison between CPI and standard plenoptic imaging. We will analyze the features of such schemes in detail, and discuss on the advantages and drawbacks related to each method.

\section{Second-order imaging with chaotic light}

The working principle of the setups that will be considered in the following sections is to perform second-order correlation imaging by splitting light from a single source in two arms, and to collect the signal at the end of the two paths by means of two sensors, $D_a$ and $D_b$.

The second-order imaging capabilities of light from chaotic and thermal sources are related with their peculiar spatio-temporal correlations. Such sources are characterized by negligible transverse coherence of the emitted light, and by large intensity fluctuations, of the same order as the time-averaged intensity. Due to these properties, the fluctuation-fluctuation correlations do not vanish in the limit of large number of photons \cite{mandel}. To unveil the correlation properties hidden in intensity fluctuations (whose typical duration coincides with the longitudinal coherence time of the source), we shall use a classical scalar model of monochromatic light with frequency $\omega$, in which the electric field on a planar source is represented by the complex random variable $E_0(\bm{\rho})$, to which intensity is related by $I_0(\bm{\rho})=|E_0(\bm{\rho})|^2$. To characterize chaotic light, we will assume that, due to random phases, the expectation values $\langle E_0(\bm{\rho}) \rangle$ of the field identically vanish,
as well as all the averages of products of an odd number of fields. The expectation value does not generally vanish, instead, if one considers products of an equal number of $E_0$ and $E_0^*$, including intensity. In the limit in which the transverse coherence area is small with respect to the extent of the intensity profile and with the assumption that the source is perfectly chaotic, one can approximate the two-point correlation function according to the Schell model \cite{mandel} as
\begin{equation}
\langle E_0(\bm{\rho}_1) E_0(\bm{\rho}_2)^* \rangle = \sqrt{ \I_0(\bm{\rho}_1) \I_0(\bm{\rho}_2) } A_c \delta^{(2)} (\bm{\rho}_1-\bm{\rho}_2),
\end{equation}
where $\I_0 := \langle I_0 \rangle$ is the average intensity, $A_c$ the transverse coherence area and $\delta^{(2)}$ the two-dimensional Dirac delta function. To estimate average intensity correlations, we will combine such properties with a Gaussian assumption on the four-point field correlations:
\begin{align}\label{4pointgauss}
& \langle E_0(\bm{\rho}_1) E_0(\bm{\rho}_2)^* E_0(\bm{\rho}_3) E_0(\bm{\rho}_4)^* \rangle \nonumber \\ & = \langle E_0(\bm{\rho}_1) E_0(\bm{\rho}_2)^* \rangle \langle E_0(\bm{\rho}_3) E_0(\bm{\rho}_4)^* \rangle \nonumber \\ & + \langle E_0(\bm{\rho}_1) E_0(\bm{\rho}_4)^* \rangle \langle E_0(\bm{\rho}_2)^* E_0(\bm{\rho}_3) \rangle.
\end{align}
This assumption respects the symmetry prescription for the two-photon expectation values. As we shall shortly see, the presence of the second term guarantees the nontrivial encoding of information in the correlation of intensity fluctuations.

The field on each detector is given by the paraxial propagators, $g_a$ and $g_b$:
\begin{equation}
E_{a,b}(\bm{\rho}) = \int \d^2\bm{\rho}_s g_{a,b}(\bm{\rho},\bm{\rho}_s) E_0(\bm{\rho}_s).
\end{equation}
This property, together with \eqref{4pointgauss} and the vanishing averages of field products containing a different number of $E$ and $E^*$, fully characterizes our model of chaotic light. By indicating with $E_a$ and $E_b$ the fields propagated from the source to detectors $D_a$ and $D_b$, respectively, and with $I_{a,b}= |E_{a,b}|^2$ the related intensities, one can combine the paraxial propagation with the chaotic nature of the source and the assumption in Eq.~\eqref{4pointgauss}, to obtain the expectation value of intensity correlations
\begin{equation}\label{intensityexp}
\langle I_a(\bm{\rho}_a) I_b(\bm{\rho}_b) \rangle = \langle I_a(\bm{\rho}_a) \rangle \langle I_b(\bm{\rho}_b) \rangle + A_c^2 \Gamma(\bm{\rho}_a,\bm{\rho}_b),
\end{equation}
where
\begin{align}
\langle I_{a,b}(\bm{\rho}) \rangle & = A_c \int \d^2\bm{\rho}_s |g_{a,b}(\bm{\rho},\bm{\rho}_s)|^2 \I_0(\bm{\rho}_s) , \label{Iab} \\
\Gamma(\bm{\rho}_a,\bm{\rho}_b) & = \left| \int \d^2\bm{\rho}_s g_a(\bm{\rho}_a,\bm{\rho}_s)^* g_b(\bm{\rho}_b,\bm{\rho}_s) \I_0(\bm{\rho}_s) \right|^2 . \label{Gammageneral}
\end{align}
The first term in Eq.\ \eqref{intensityexp} represents the mere product of the average intensities measured at the two detectors, and does not contain any information that cannot be retrieved in a first-order measurement. The nontrivial part of the correlation function, $\Gamma(\bm{\rho}_a,\bm{\rho}_b)$, that coincides up to $A_c^2$ with the \textit{correlation of intensity fluctuations}, contains information that is not encoded in the first-order measurement of intensities. Intuitively, if the integrand $g_a^*g_b$ in Eq.~\eqref{Gammageneral} is concentrated around a single point $\bm{\rho}_s$ on the source plane, the measurement of $\Gamma$ will simply give information on the intensity distribution on the source; most interesting, if an object is placed in one of the two arms, an image of such object can be retrieved through correlation measurements between $D_a$ and $D_b$. This principle provides the basis of ghost imaging with chaotic light \cite{valencia,scarcelliPRL}, which, however, does not fully exploits the four-dimensional nature of $\Gamma(\bm{\rho}_a,\bm{\rho}_b)$. In the following section, we shall present different schemes to encode a \textit{double} imaging effect in the correlation function, thus enabling to perform the typical tasks of plenoptic imaging, such as refocusing and changing the point of view on the scene, as required for 3D imaging.

Experimentally, the intensity correlation can be retrieved by sampling intensities at the end of each arm by synchronized cameras. The product of each pair of frames (one from each camera) will be time-averaged to get the expectation value appearing in Eq.\ \eqref{intensityexp}. In order to directly retrieve the interesting part of the correlation [Eq. \eqref{Gammageneral}], the constant average value of the intensity measured by each sensor can be electronically removed by applying a DC block (or high-pass filter) before correlating the signals. To ensure the effectiveness of the procedure, the detector must have a response time similar to the typical timescale of intensity fluctuations, as determined by the coherence time of the source.

\section{Plenoptic imaging by correlation of first-order images - scheme I}

\begin{figure}
\centering
\includegraphics[width=0.48\textwidth]{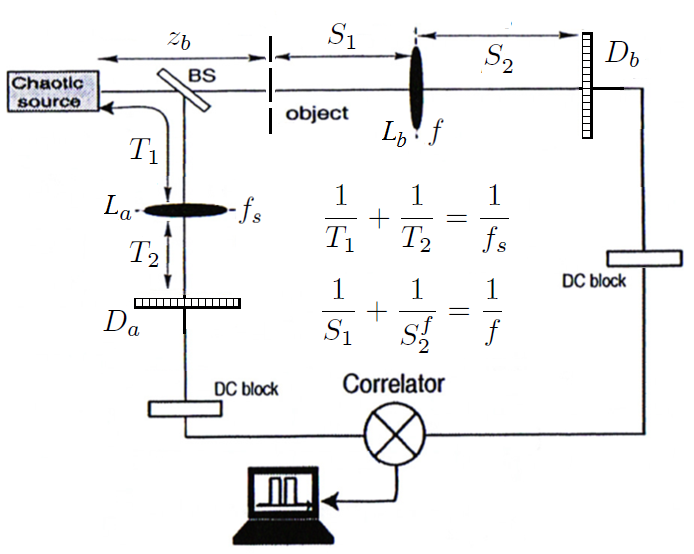}
\caption{Scheme I. Light from a chaotic source is separated in two arms by a beam-splitter (BS) and sent to two distant sensors $D_a$ and $D_b$. First-order imaging of both the object and the source are performed in the transmitted and reflected arm of the BS, respectively, by means of two positive lenses of focal length $f$ and $f_s$. By correlating the intensity fluctuations measured by the two detectors, we retrieve the combined information on the two images, as encoded in $\Gamma(\bm{\rho}_a,\bm{\rho}_b)$, in the attempt to reconstruct the path of light in the setup, and thus enable refocusing of blurred images retrieved whenever $S_2\neq S_2^f$. The DC blocks are high-pass filters that cut the average intensities measured by the two sensors [see Eq.\ \eqref{intensityexp}], thus sending to the correlator only the intensity fluctuations $\Delta I_{a,b}=I_{a,b}-\langle I_{a,b} \rangle$. }\label{fig.PCC}
\end{figure}

Plenoptic imaging relies on the possibility of recording the direction of light propagating in the device. An intuitive way to achieve this goal is to attempt exploiting the peculiar correlations of chaotic light by measuring the intensity correlations between the image of the scene and the image of the source, as recorded by two separated sensors. The underpinning idea is to track down the light paths connecting the source to the object. To implement this scheme, as shown in Figure~\ref{fig.PCC}, light from a chaotic source is split in two paths, and a positive lens is placed along each one. A lens ($L_a$) with focal length $f_s$, is located in the transmitted path at distances $T_1$ from the source and $T_2$ from the sensor $D_a$, such that
\begin{equation}\label{fs}
\frac{1}{T_1} + \frac{1}{T_2} = \frac{1}{f_s};
\end{equation}
the image of the source is thus retrieved by the sensor $D_a$. In the reflected path, a planar transmitting object is placed at an arbitrary distance $z_b$ from the source, and a positive lens ($L_b$) with focal length $f$ is located at a distance $S_1$ from it, and reproduces its image on the sensor $D_b$. In order to prove the refocusing capability of this scheme, we will admit that the distance $S_2$ between $L_b$ and $D_b$ can differ (by a positive or negative value) from the value $S_2^f$ that ensures focusing the image of the object on $D_a$, which is
\begin{equation}\label{fo}
\frac{1}{S_1} + \frac{1}{S_2^f} = \frac{1}{f}.
\end{equation}

To compute the correlation of intensity fluctuations [Eq.~\eqref{Gammageneral}] and unveil its plenoptic properties, let us start by evaluating the transfer functions of the two arms, $g_a$ and $g_b$, encoding the first-order image of the source and the object, respectively. In the transmitted arm, the propagator reads
\begin{align}\label{gafirst}
g_a(\bm{\rho}_a,\bm{\rho}_s) & = \int \d^2\bm{\rho}_{\ell_s} h(\bm{\rho}_a-\bm{\rho}_{\ell_s},T_2) P_s( \bm{\rho}_{\ell_s} ) \e^{-\frac{\i k}{2f_s}\bm{\rho}_{\ell_s}^2} \nonumber \\
& \quad \times h(\bm{\rho}_{\ell_s}-\bm{\rho}_s,T_1) \nonumber \\
& = -\frac{k^2 \e^{\i k \!\left[T_1+T_2 +\frac{1}{2}\!\left(\frac{\bm{\rho}_s^2}{T_1}+ \frac{\bm{\rho}_a^2}{T_2}\right)\right]} \tilde{P}_s \!\left[\frac{k}{T_1}\!\left( \bm{\rho}_s + \frac{\bm{\rho}_a}{M_s} \right)\!\right]\! }{(2\pi)^2 T_1 T_2} ,
\end{align}
where $P_s$ is the pupil function of lens $L_b$, $\tilde{P}_s(\bm{q})=\int \d^2\bm{\rho} P_s(\bm{\rho}) \e^{-\i\bm{\rho}\cdot\bm{q}}$ its Fourier transform, and $M_s=T_2/T_1$ the lens magnification.
In this equation, we have introduced the paraxial propagator
\begin{equation}
h(\bm{\rho},z) = -\frac{\i k}{2\pi z} \exp\left( \i k z + \frac{\i k}{2z} \bm{\rho}^2 \right),
\end{equation}
with $k=\omega/c=2\pi/\lambda$ the longitudinal component of the wave vector, that will appear extensively throughout the paper. The result \eqref{gafirst}, which is particularly simple due to the focusing condition of Eq.~\eqref{fs}, entails the image of the source intensity profile in the first-order intensity defined in Eq.~\eqref{Iab}: 
\begin{equation}
\langle I_a(\bm{\rho}_a) \rangle \propto \int \d^2\bm{\rho}_s \I_0(\bm{\rho}_s) \!\left| \tilde{P}_s \!\left[\frac{k}{T_1}\!\left( \bm{\rho}_s + \frac{\bm{\rho}_a}{M_s} \right)\!\right]\!  \right|^2
\end{equation}
The propagator in the reflected arm can be computed in a similar way, but taking into account the presence of the object, characterized by the transmission function $A(\bm{\rho}_o)$, and placed at a distance $z_b$ from the source:
\begin{align}
g_b(\bm{\rho}_b,\bm{\rho}_s) & = \int \d^2\bm{\rho}_{\ell} h(\bm{\rho}_b-\bm{\rho}_{\ell},S_2) P_s( \bm{\rho}_{\ell} ) \e^{-\frac{\i k}{2f}\bm{\rho}_{\ell}^2} \nonumber \\
& \times\!\int\!\d^2\bm{\rho}_o h(\bm{\rho}_{\ell}-\bm{\rho}_o,S_1) A(\bm{\rho}_o) h(\bm{\rho}_o-\bm{\rho}_s,z_b) \nonumber \\
& = \frac{i k^3 \e^{\i k\!\left[ z_b+S_1+S_2 + \frac{1}{2}\!\left(\frac{\bm{\rho}_s^2}{z_b}+ \frac{\bm{\rho}_b^2}{S_2}\right)\!\right]} }{(2\pi)^3 z_b S_1 S_2} \nonumber \\
& \times \!\int\!\d^2\bm{\rho}_o\d^2\bm{\rho}_{\ell} A(\bm{\rho}_o) P (\bm{\rho}_{\ell}) \e^{\i k \bigl( \phi(\bm{\rho}_o,\bm{\rho}_{\ell},\bm{\rho}_b) - \frac{\bm{\rho}_o\cdot\bm{\rho}_s}{z_b}\bigr)} ,
\end{align}
with $P$ the pupil function of lens $L_b$, and
\begin{align}\label{phasefirst}
\phi(\bm{\rho}_o,\bm{\rho}_{\ell},\bm{\rho}_b) = & \frac{\bm{\rho}_{\ell}^2}{2}\left(\frac{1}{S_2}-\frac{1}{S_2^f}\right) + \frac{\bm{\rho}_{o}^2}{2}\left(\frac{1}{z_b}+\frac{1}{S_1}\right) \nonumber \\ & - \left(\frac{\bm{\rho}_o}{S_1} + \frac{\bm{\rho}_b}{S_2}\right)\cdot \bm{\rho}_{\ell} ,
\end{align}
where $M=S_2/S_1$ is the magnification of this lens. Notice that, while quadratic terms in the lens-plane coordinate $\bm{\rho}_{\ell}$ disappear after imposing the focusing condition $S_2=S_2^f$, quadratic phases in $\bm{\rho}_o$ cannot disappear from the integral.

The function $\Gamma(\bm{\rho}_a,\bm{\rho}_b)$ can now be computed through Eq.~\eqref{Gammageneral}. Without further approximations, the correlation of intensity fluctuations is a rather complicated integral that depends on four functions: the aperture of the object, the pupils of the lenses $L_a$ and $L_b$, and the intensity profile of the source. The aperture of the lens $L_b$ uniquely determines the resolution of the first-order image of the object; however, in second-order imaging, both the source and the lens $L_a$ can act as additional pupils, selecting the directions of rays that contribute to the build-up of correlations. Let us now work in the hypothesis that the resolution of the object image is defined, even at second order, only by the aperture of the lens $L_b$. Hence, the lens $L_a$ is assumed to be diffraction-limited and the source to be so large as not to affect propagation of rays emitted by the source and passing through the lens. The correlation function $\Gamma$ encodes plenoptic information, as can be seen from its expression (up to an irrelevant constant)
\begin{align}\label{Gammafirst}
& \Gamma(\bm{\rho}_a,\bm{\rho}_b) = \nonumber \\ & \qquad \left| \!\int\!\d^2\bm{\rho}_o\d^2\bm{\rho}_{\ell} A(\bm{\rho}_o) P (\bm{\rho}_{\ell}) \e^{\i k \bigl( \phi(\bm{\rho}_o,\bm{\rho}_{\ell},\bm{\rho}_b) + \frac{\bm{\rho}_o\cdot\bm{\rho}_a}{M_s z_b} \bigr) } \right|^2 .
\end{align}
In this regime, the measurement of correlations between intensities retrieved in $\bm{\rho}_a$ and $\bm{\rho}_b$ provides a structured information on the ``points'' of both the object and the lens $L_b$ where light has passed, with a resolution fixed by the point-spread functions (PSF).
By integrating over $\bm{\rho}_a$, i.e.~by correlating each pixel of $D_b$ with the whole detector $D_a$, one recovers the first-order incoherent image of the object
\begin{align}\label{ghostfirst}
\Sigma(\bm{\rho}_a)&=\int\!\d^2\bm{\rho}_a \Gamma(\bm{\rho}_a,\bm{\rho}_b) = \Bigl(2\pi \frac{M_s z_b}{k}\Bigr)^2 \!\int\!\d^2\bm{\rho}_o |A(\bm{\rho}_o)|^2 \nonumber \\
& \quad \times \!\left|\!\int\!\d^2\bm{\rho}_{\ell} P(\bm{\rho}_{\ell}) \e^{\i k \bigl[ \bigl( \frac{1}{S_2} - \frac{1}{S_2^f} \bigr) \frac{\bm{\rho}_{\ell}^2}{2} - \bigl( \frac{\bm{\rho}_o}{S_1} + \frac{\bm{\rho}_b}{S_2}\bigr)\cdot \bm{\rho}_{\ell} \bigr] }\!\right|^2 \ .
\end{align}
This image is focused on $D_b$ when $S_2=S_2^f$, and its point-spread function is given by the Fourier transform of the lens pupil function $P$, centered on $\bm{\rho}_{b}=-(S_2^f/S_1)\bm{\rho}_o$, as expected. However, this integration inhibits plenoptic imaging since it completely erases the information on the light direction. 

In view of plenoptic imaging, we shall thus focus on the correlation function associated with a single pixel of $D_a$, such as $\bm{\rho}_a=0$. As can be shown based on Eq.~\eqref{Gammafirst}, the (coherent) image of the object is actually compromised in this case: its PSF is again determined by the lens $L_b$, but effective focusing is inhibited by the presence of the $\bm{\rho}_o$-dependent quadratic phase, which comes from free propagation between the source and the object  [see Eq.~\eqref{phasefirst}]. This can be easily seen by considering that a relevant property of the four-variable function $\Gamma$, which is fundamental for its plenoptic application, is its ability to encode the projection of the focusing element (here, the lens $L_b$), which enables tracing the light paths from the object to the focusing element itself. In this perspective, it is worth analyzing the correlation function corresponding to a point-like lens $P(\bm{\rho}_{\ell})=\delta^{(2)}(\bm{\rho}_{\ell}-\bar{\bm{\rho}}_{\ell})$, which reads
\begin{align}
& \Gamma(\bm{\rho}_a,\bm{\rho}_b)\bigl|_{\mathrm{p.l.}} = \nonumber \\
& \qquad \!\left| \!\int\!\d^2\bm{\rho}_o A(\bm{\rho}_o) \e^{\i k \bigl[ \bigl( \frac{1}{S_2} - \frac{1}{S_2^f} \bigr) \frac{\bm{\rho}_o^2}{2} + \bm{\rho}_o\cdot \bigl(\frac{\bm{\rho}_a}{M_s z_b} - \frac{\bar{\bm{\rho}}_{\ell}}{S_1} \bigr)\bigr] } \!\right|^2\!.
\end{align}  
Due to the quadratic phase factor, this correlation function does not properly reproduces the image of the chosen point on the lens plane. As we shall see shortly, this additional phase produces, in the geometrical-optics limit, a $\bm{\rho}_b$-dependent shift in the image of the lens, which significantly complicates the refocusing algorithm. All the issues related with the presence of such undesired term in the phase will be overcome by the next two schemes, which exploit the less intuitive but more effective ghost imaging phenomenon \cite{laserphys,valencia,scarcelliPRL} to deeply exploit the intrinsic position-momentum correlation of chaotic light.

To unveil the plenoptic properties of the present correlation function, we shall consider the geometrical-optics limit (i.e., $\omega\to\infty$): in this case, the most prominent contribution to the integral in Eq.~\eqref{Gammafirst} reads 
\begin{equation}\label{Gammastatfirst}
\Gamma(\bm{\rho}_a,\bm{\rho}_b) \sim \left|A(\bar{\bm{\rho}}_o(\bm{\rho}_a,\bm{\rho}_b))\right|^2 \left|P(\bar{\bm{\rho}}_{\ell}(\bm{\rho}_a,\bm{\rho}_b))\right|^2 ,
\end{equation}
where the arguments $\bar{\bm{\rho}}_o$ and $\bar{\bm{\rho}}_{\ell}$ are the stationary points of the phase appearing under the integral in Eq.\ \eqref{Gammafirst}. In the focused case, $\bar{\bm{\rho}}_o= -(S_1/S_2^f)\bm{\rho}_b$, as expected, and integration over $\bm{\rho}_a$ gives the standard image: $\Sigma(\bm{\rho}_b;S_f)\sim|A(-(S_1/S_2^f)\bm{\rho}_b)|^2$. If the object is out of focus, $\bar{\bm{\rho}}_o$ becomes dependent on \textit{both} $\bm{\rho}_b$ and $\bm{\rho}_a$. Hence, if one integrates over $\bm{\rho}_a$ as in Eq.\ \eqref{ghostfirst} to increase the signal-to-noise ratio and obtain an incoherent image, the quasi one-to-one correspondence $\bm{\rho}_o \leftrightarrow -(S_1/S_2^f)\bm{\rho}_b$ is lost and the image appears blurred. However, retrieving the whole function $\Gamma(\bm{\rho}_a,\bm{\rho}_b)$ provides access to \textit{both} $\bar{\bm{\rho}}_o$ and $\bar{\bm{\rho}}_{\ell}$; in this case, the blurred image can be refocused by applying the following linear scaling and  proper combination of the detector coordinates:
\begin{equation}
\Sigma_{\mathrm{ref}} (S_2) = \int\d^2\bm{\rho}_a \Gamma(\bm{\rho}'_a,\bm{\rho}'_b) \sim \left| A\left( - \frac{S_1}{S_2^f} \bm{\rho}_b \right) \right|^2 ,
\end{equation}
with
\begin{align}
\bm{\rho}'_a = & \eta(S_2) \left( 1 + \frac{S_1(S_2^f-S_2)}{S_2 S_2^f} \right)^{-1} \bm{\rho}_a , \\
\bm{\rho}'_b = & \eta(S_2) \left[ \frac{S_2}{S_2^f} \bm{\rho}_b - \left( 1+ \frac{S_2 S_2^f}{S_1(S_2^f-S_2)} \right)^{-1} \!\frac{S_2}{M_s z_b} \bm{\rho}_a \right] .
\end{align}
with
\begin{equation}
\eta(S_2)= 1- \frac{S_1(S_2^f-S_2)}{S_2 S_2^f} \left( 1+ \frac{S_1}{z_b} \right) .
\end{equation} 
Such formulas represent the generalization to second-order imaging of the refocusing algorithm typical of standard plenoptic cameras \cite{ng}. Notice, however, that in this case \textit{both} sensor coordinates must be transformed.

As we shall see in the next sections, the ghost imaging phenomenon enables to design correlation plenoptic imaging protocols characterized by the decoupling of the directional and the spatial detection; in this case, pixels on the directional detector are \textit{uniquely} associated to points on the focusing element, thus highly simplifying the refocusing algorithm. 

\section{Plenoptic imaging by correlation of ordinary and ghost images - schemes II and III}

In the previous section, we have demonstrated how to perform plenoptic imaging by correlating two first-order images. Such a procedure leads to a quite complicate refocusing algorithm due to the presence of detrimental phase factors, stemming from the fact that the setup does not produce a real image of the main focusing element, namely the lens that focuses the object.  The problem can be overcome by fully exploiting the spatio-temporal correlation properties of chaotic light, namely, by implementing the \textit{ghost imaging} protocol \cite{pittman,gatti,laserphys,valencia,scarcelliPRL} so as to retrieve a purely second-order image in the absence of its first order counterpart. In the lensless configuration of ghost imaging \cite{scarcelliPRL}, the object is placed in one arm and a detector with no spatial resolution is placed just behind it in order to collect all the transmitted light; correlations are measured between this detector and a spatially-resolving detector, which is placed in the other arm. When the object and the spatially-resolving detector are at the same distance from the source, a real stigmatic image of the object is retrieved by second-order correlation measurement. In this case, the source acts as a focusing element: its transverse width and longitudinal distance from the detector determine the intrinsic resolution and depth of field of the ghost image.

\begin{figure}
\centering
\includegraphics[width=0.48\textwidth]{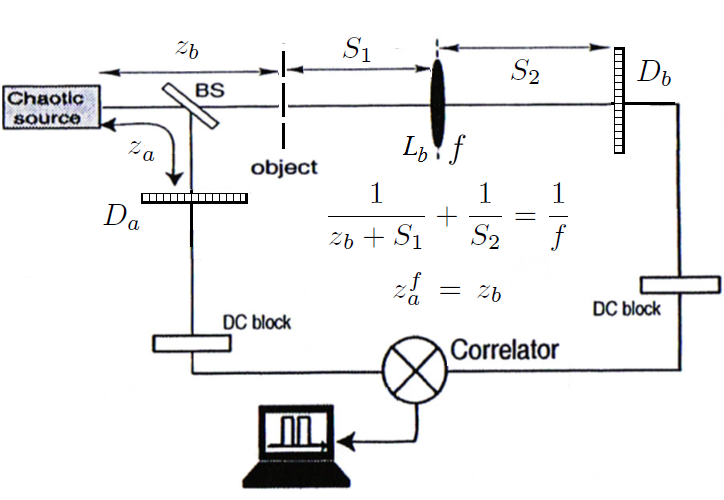}
\caption{\small Scheme II. Experimental setup proposed in \cite{cpi_prl}, in which the ghost image of the object, illuminated by chaotic light, is retrieved on the sensor $D_a$ by second-order correlation measurements of this whole sensor with each pixel of the sensor $D_b$. The lens in the transmitted arm of the beam splitter (BS) serves at reproducing, on the sensor behind it, a first-order image of the focusing element, namely, the source. The point on the source, from which the signal is emitted, is thus correlated with the point of the object by which it is transmitted. Plenoptic information retrieved by correlation measurements between the intensity fluctuations measured by $D_a$ and $D_b$ (which are cleaned of the average intensities by the DC blocks) enables refocusing images when $z_a\neq z_b$.  }\label{fig.PCI}
\end{figure}

In \cite{cpi_prl}, we proposed a modified ghost imaging scheme in which both detectors are spatially resolving. Here, we review the features and plenoptic properties of such setup, and propose an alternative ghost-imaging-based configuration. In the setup represented in Figure 2, light transmitted by the beam splitter encounters first an object at a distance $z_b$, characterized by the aperture $A(\bm{\rho}_o)$, and then a positive lens $L_b$, before impinging on the sensor $D_b$. In the reflected arm, light propagates freely on a distance $z_a$ from the source to the detector $D_a$, as in the lensless ghost imaging scheme. The position of the lens is chosen to image the source on $D_b$, hence,
\begin{equation}
\frac{1}{z_b+S_1}+\frac{1}{S_2} = \frac{1}{f},
\end{equation}
where $S_1$ and $S_2$ are the object-to-lens and lens-to-sensor distances, respectively, and $f$ is the focal length of $L_b$. Computation of the propagators and integration over the source as in \eqref{Gammageneral} yield the result \cite{cpi_prl,cpi_qmqm}
\begin{align}\label{Gammasecond}
\Gamma(\bm{\rho}_a,\bm{\rho}_b) = & \Biggl| \int\!\d^2\bm{\rho}_o \d^2\bm{\rho}_s  A(\bm{\rho}_0) \mathcal{I}_0 (\bm{\rho}_s) \nonumber \\
& \e^{\i k \bigl( \psi_1(\bm{\rho}_o,\bm{\rho}_s,\bm{\rho}_a) - \frac{\bm{\rho}_s\cdot\bm{\rho}_b}{M z_b} \bigr)} \Biggr|^2,
\end{align}
where
\begin{align}\label{psisecond}
\psi_1(\bm{\rho}_o,\bm{\rho}_s,\bm{\rho}_a) = \left( \frac{1}{z_b} - \frac{1}{z_a} \right) \frac{\bm{\rho}_s^2}{2} - \left( \frac{\bm{\rho}_o}{z_b}  - \frac{\bm{\rho}_a}{z_a} \right) \cdot \bm{\rho}_s  ,
\end{align}
with $M=S_2/(S_1+z_b)$ the magnification of the source on the $D_b$. In the above equations, we have assumed the lens $L_b$ to be diffraction limited. The integral over $\bm{\rho}_b$ yields the incoherent (ghost) image of the object 
\begin{align}
\Sigma(\bm{\rho}_a) = & \int\!\d^2\bm{\rho}_b \Gamma(\bm{\rho}_a,\bm{\rho}_b) = \Bigl(2\pi \frac{M z_b}{k}\Bigr)^2 \!\int\!\d^2\bm{\rho}_o |A(\bm{\rho}_o)|^2 \nonumber \\
& \times \left| \int\d^2\bm{\rho}_s \mathcal{I}_0(\bm{\rho}_s) \e^{\i k \psi_1(\bm{\rho}_o,\bm{\rho}_s,\bm{\rho}_a)} \right|^2
\end{align}
that is focused when $z_a=z_a^f=z_b$, and whose point-spread function is determined by the Fourier transform $\tilde{\mathcal{I}}_0(k(\bm{\rho}_o-\bm{\rho}_a)/z_a^f)$ of the source. In the focused case, it is evident that the only quadratic phase in \eqref{Gammasecond}-\eqref{psisecond} disappears. The resolution of the focused ghost image can be estimated by $\Delta \bm{\rho}_a \sim \lambda z_a^f/D_s$, with $D_s$ the effective diameter of the source. On the other hand, one can check that the correlation function $\Gamma$ keeps track of the first-order imaging of the source. A coherent image of the source intensity profile can be read on $D_b$, in correspondence of each fixed point on $D_a$, and integration over $D_a$ provides the incoherent image of the (squared) intensity profile, with point-spread function of width $\Delta \bm{\rho}_s \sim \lambda z_b/a$, where $a$ is the smallest length scale of the object details. The resolution on the source is thus determined by diffraction at the object.

The behavior in the geometrical optics limit can be again determined by the stationary-phase approximation, providing
\begin{equation}
\Gamma(\bm{\rho}_a,\bm{\rho}_b) \sim \mathcal{I}_0 \!\left(\! -\frac{\bm{\rho}_b}{M} \!\right)^2\! \left| A \!\left[ \frac{z_b}{z_a}\bm{\rho}_a - \frac{\bm{\rho}_b}{M} \!\left(1-\frac{z_b}{z_a}\right)\!\right]\! \right|^2 .
\end{equation}
Unlike the previous case, each point on the focusing element uniquely corresponds to a point on the sensor $D_b$. As for the focused image, it is reproduced with unit magnification on $D_a$, and remains unaffected by integration of $\bm{\rho}_b$. Instead, when $z_a\neq z_b$, the additional dependence on $\bm{\rho}_b$ induces a loss of information when integrating the signal on $D_b$. However, if no integration is performed, and the whole $\Gamma(\bm{\rho}_a,\bm{\rho}_b)$ is retrieved, the dependence on $\bm{\rho}_b$ becomes a resource, providing different points of view on the out-of-focus object.
Moreover, unlike in the first setup [see Eq.~\eqref{Gammastatfirst}] the focused incoherent image can be reconstructed by shifting \textit{just one 2d coordinate}:
\begin{equation}
\Sigma_{\mathrm{ref}}(\bm{\rho}_a)\! = \!\int\d^2\!\bm{\rho}_b \Gamma\!\left[\! \frac{z_a}{z_b}\bm{\rho}_a - \frac{\bm{\rho}_b}{M} \!\left(1-\frac{z_a}{z_b}\right), \bm{\rho}_b \!\right] \sim |A(\bm{\rho}_a)|^2 .
\end{equation} 
The effectiveness of such procedure has been experimentally demonstrated in \cite{cpi_exp}. We postpone a comparison with the method discussed in the previous paragraph after presentation of the third protocol.

\begin{figure}
\centering
\includegraphics[width=0.48\textwidth]{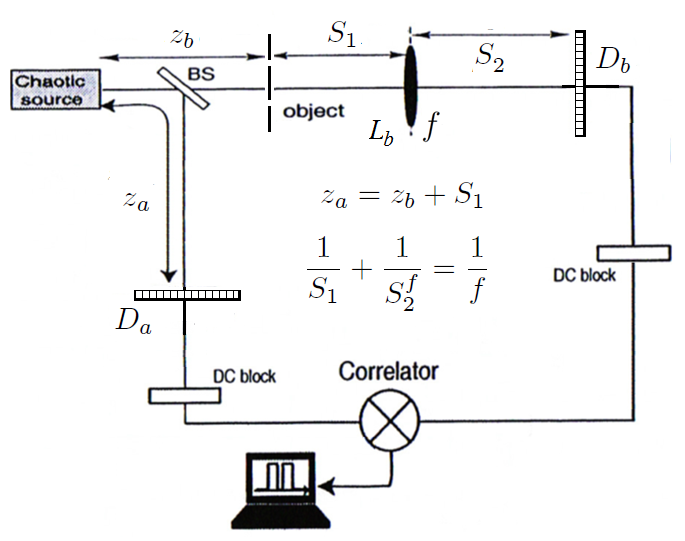}
\caption{\small Scheme III. In the transmitted arm of the BS, the first-order image of the object is reproduced by the lens $L_b$  on the sensor $D_b$; in the reflected arm, the ghost image of this lens is reproduced on sensor $D_a$, thanks to the chaotic nature of the light source and to correlation measurements. The plenoptic properties characterizing the correlation of intensity fluctuations enable refocusing objects placed in $S_2\neq S_2^f$. }\label{fig.PCI2}
\end{figure}

We are now going to uncover the possibility of performing plenoptic imaging, hence refocusing, in the setup represented in Figure \ref{fig.PCI2}, characterized by the same components as scheme II, but with the light source and the lens playing opposite roles. In fact, in this case, the lens $L_b$ \textit{focuses the object}, at first-order, on the detector $D_b$, while the chaotic source and correation measurements enable reproducing a \textit{ghost image of the lens} on $D_a$. To ensure focusing of the latter, the relation
\begin{equation}\label{focusingthird}
z_a=z_b+S_1
\end{equation}
must hold. The object is focused on $D_b$ when the lens-to-sensor distance is $S_2=S_2^f$, with
\begin{equation}
\frac{1}{S_1}+\frac{1}{S_2^f} = \frac{1}{f};
\end{equation}
however, for studying the refocusing power of this scheme, we shall not fix the value of $S_2$. Computation of the propagator from the source to $D_a$ is immediate
\begin{equation}
g_a(\bm{\rho}_a,\bm{\rho}_s) = h(\bm{\rho}_a-\bm{\rho_s},z_a) = -\frac{\i k \e^{\i k \bigl[ z_a + \frac{1}{z_a}\bigl(\frac{\bm{\rho}_a^2+\bm{\rho}_s^2}{2} - \bm{\rho}_a\cdot\bm{\rho}_s \bigr)\bigr]} }{2\pi z_a}  ,
\end{equation}
while the propagator in the transmitted arm reads
\begin{align}
g_b(\bm{\rho}_b,\bm{\rho}_s) & \!=\! \int\!\d^2\!\bm{\rho}_o\! \int\!\d^2\!\bm{\rho}_{\ell} h(\bm{\rho}_b-\bm{\rho}_{\ell},S_2) P(\bm{\rho}_{\ell}) \e^{-\frac{\i k}{2 f}\bm{\rho}_{\ell}^2} \nonumber \\
& \times h(\bm{\rho}_{\ell}-\bm{\rho}_o) A(\bm{\rho}_o) h(\bm{\rho}_o-\bm{\rho}_s) \nonumber \\
& =  \frac{\i k^3 \e^{\i k \bigl[ z_b+S_1+S_2 + \frac{1}{2}\bigl(\frac{\bm{\rho}_b^2}{S_2}+\frac{\bm{\rho}_s^2}{z_b} \bigr) \bigr]} }{2\pi z_b S_1 S_2} \! \int\!\d^2\!\bm{\rho}_o\! \int\!\d^2\!\bm{\rho}_{\ell} \nonumber \\
& \times A(\bm{\rho}_o) P(\bm{\rho}_{\ell})  \e^{\i k \bigl[ \psi(\bm{\rho}_o,\bm{\rho}_{\ell},\bm{\rho}_b) + \bigl( \frac{1}{z_b}+\frac{1}{S_1} \bigr) \frac{\bm{\rho}_o^2}{2} - \frac{\bm{\rho}_o\cdot\bm{\rho}_s}{z_b} \bigr] } ,
\end{align}
with
\begin{equation}\label{psithird}
\psi_2(\bm{\rho}_o,\bm{\rho}_{\ell},\bm{\rho}_b) = \left(\frac{1}{S_2}-\frac{1}{S_2^f}\right) \frac{\bm{\rho}_{\ell}^2}{2} - \left( \frac{\bm{\rho}_o}{S_1} + \frac{\bm{\rho}_b}{S_2} \right) \cdot \bm{\rho}_{\ell}.
\end{equation}
The evaluation of the correlation in Eq. \eqref{Gammageneral} is complicated by the presence of integrals over three functions: the source intensity profile, the lens pupil function and the object aperture function. However, also in this case, we shall assume that the ranges of $\bm{\rho}_a$ and $\bm{\rho}_b$ over which $\Gamma$ is non-vanishing are determined by the sizes of the object and of its focusing element $L_b$; the source will thus be considered to be so large as not to affect the propagation of light from the source to the detection plane. Under such approximation, we obtain an expression similar to the one in Eq. \eqref{Gammasecond}, namely (up to irrelevant constants),
\begin{equation}\label{Gammathird}
\Gamma(\bm{\rho}_a,\bm{\rho}_b) \!=\! \left|\!\int\!\d^2\bm{\rho}_o \d^2\bm{\rho}_{\ell}  A(\bm{\rho}_0) P(\bm{\rho}_s) \e^{\i k \!\bigl(\! \psi_2(\bm{\rho}_o,\bm{\rho}_{\ell},\bm{\rho}_b) + \frac{\bm{\rho}_o\cdot\bm{\rho}_a}{S_1} \!\bigr)} \!\right|^2,
\end{equation}
with the source intensity profile replaced by the pupil function of the lens, and the focusing condition $z_a=z_b$ replaced by $S_2=S_2^f$. Similar to previous schemes, the incoherent image of the object is recovered by summing the correlation function over $D_a$
\begin{align}
\Sigma(\bm{\rho}_b;S_2) = & \int\!\d^2\bm{\rho}_a \Gamma(\bm{\rho}_a,\bm{\rho}_b) = \!\left(\!2\pi \frac{S_1}{k}\!\right)^2 \!\int\!\d^2\bm{\rho}_o |A(\bm{\rho}_o)|^2 \nonumber \\
& \times \left| \int\d^2\bm{\rho}_s P(\bm{\rho}_{\ell}) \e^{\i k \psi_2(\bm{\rho}_o,\bm{\rho}_{\ell},\bm{\rho}_b)} \right|^2 ,
\end{align}
where it is evident that the focusing condition is related with the disappearance of the only quadratic phase in Eq. \eqref{psithird}. As expected, the width of the point-spread function is determined by the lens diameter $D_{\ell}$, which is, $\Delta\rho_b \sim \lambda S_2^f/D_{\ell}$; hence, the object resolution is $\Delta\rho_o\sim \lambda S_1/D_{\ell}$. On the other hand, the incoherent ghost image of the lens can be retrieved by integrating the correlation function over $\bm{\rho}_a$
\begin{equation}
\int\!\d^2\bm{\rho}_b \Gamma(\bm{\rho}_a,\bm{\rho}_b) \propto \int\!\d^2\bm{\rho}_{\ell} |P(\bm{\rho}_{\ell})|^2 \left| \tilde{A} \!\left[ \frac{k}{S_1} (\bm{\rho}_{\ell} - \bm{\rho}_a)  \right]\right|^2;
\end{equation}
its point-spread function is determined by the Fourier transform of the object aperture, yielding the resolution $\Delta\rho_a\sim \lambda S_1/a$, with $a$ the size of the smallest object details. Once again, directional resolution is hindered only by diffraction at the object, provided the source is sufficiently large.

Let us now move to the plenoptic properties of the correlation function associated with this new setup. The stationary phase under the integral in \eqref{Gammathird} indicates that the dominant contribution in the geometrical optics limit reads
\begin{equation}
\Gamma(\bm{\rho}_a,\bm{\rho}_b) \sim |P(\bm{\rho}_a)|^2 \left| A \left[ \frac{S_1}{S_2} \!\left( -\bm{\rho}_b + \!\left( 1- \frac{S_2}{S_2^f} \right)\!\bm{\rho}_a \right)\! \right] \right|^2.
\end{equation}
Also in this case, the focusing condition yields the dependence of the image on just one coordinate, which enables to keep a quasi one-to-one correspondence after integrating over $\bm{\rho}_a$. As in the previous setup, the dependence on the focusing element coordinates provides different viewpoints on the object. If the sensor $D_b$ is displaced with respect to the focusing distance $S_2^f$, the structure of the correlation function enables to perform refocusing with the following algorithm
\begin{align}
\Sigma_{\mathrm{ref}}(\bm{\rho}_b)\! & = \!\int\d^2\!\bm{\rho}_a \Gamma\!\left[ \bm{\rho}_a, \frac{S_2}{S_2^f}\bm{\rho}_b + \!\left(1-\frac{S_2}{S_2^f}\right) \bm{\rho}_a \right] \nonumber \\ & \sim \left| A \left(-\frac{S_1}{S_2^f}\bm{\rho}_a \right) \right|^2 ,
\end{align} 
which basically shifts the variable $\bm{\rho}_b$, and yields an approximation of the focused image with an accuracy that improves as the geometrical-optics limit is approached.

\begin{table}[]
\centering
\label{my-label}
\begin{tabular}{clclcll}
\hline
\textbf{\begin{tabular}[c]{@{}c@{}}Setup\\ (figure)\end{tabular}} &  & \textbf{\begin{tabular}[c]{@{}c@{}}Focusing\\ element\end{tabular}} &  & \textbf{\begin{tabular}[c]{@{}c@{}}Image of\\ focusing el.\end{tabular}} &  & \multicolumn{1}{c}{\textbf{Conditions}}                                                                 \\ \hline
\multicolumn{1}{l}{}                                              &  & \multicolumn{1}{l}{}                                                &  & \multicolumn{1}{l}{}                                                     &  &                                                                                                         \\
1                                                                 &  & Lens $L_b$                                                          &  & No                                                                       &  & \multicolumn{1}{c}{\begin{tabular}[c]{@{}c@{}}Large source\\ Diffraction-limited $L_a$\end{tabular}} \\
\multicolumn{1}{l}{}                                              &  & \multicolumn{1}{l}{}                                                &  & \multicolumn{1}{l}{}                                                     &  &                                                                                                         \\
2                                                                 &  & Source                                                              &  & Yes                                                                      &  & Diffraction-limited $L_b$                                                                               \\
\multicolumn{1}{l}{}                                              &  & \multicolumn{1}{l}{}                                                &  & \multicolumn{1}{l}{}                                                     &  &                                                                                                         \\
3                                                                 &  & Lens $L_b$                                                          &  & Yes                                                                      &  & \multicolumn{1}{c}{Large source}                                                                     \\ \hline
\end{tabular}
\caption{Comparative table of the three schemes discussed in the paper. The setup referring to Figure 1 is the only one that does not exploit ghost imaging. By ``large'' source it is meant that the source transverse size does not significantly affect propagation of rays that pass through the object and the lens $L_b$.}\label{table}
\end{table}

\section{Comparison and conclusions}

We have investigate the plenoptic capability of the intensity correlation function retrieved in three different setups, sharing a transmissive object illuminated by a chaotic light source. The main features of the three schemes are reported in Table 1. The first setup does not fully exploit the position-momentum correlation of a chaotic light source; this reflects into the impossibility to properly image the focusing element, which also leads to a complicate refocusing algorithm. On the other hand, the two plenoptic procedures relying on ghost imaging are strictly equivalent in their concept and basic setup; the difference in the role played by the optical components can lead to a preference for one or the other, based on purely practical reasons. The latter setup, in which the object is imaged by a lens, at first order, ensures a larger control on the image resolution, which is solely defined by the lens diameter (as opposed to the intensity profile of the chaotic source, as in the second scheme). Another advantage of the third scheme is that, here, the correlation between intensity fluctuations $\Gamma$ depends on the square of the pupil function $P(\bm{\rho}_{\ell})$, that usually takes values of either $0$ or $1$; on the contrary, the dependence on the squared intensity, characterizing scheme II, leads to a reduction of the signal in regions of lower intensities. On the other hand, scheme III has the disadvantage of requiring a source that is so large as not to affect the resolution of the ghost image of the lens. An interesting follow-up of this research is the application of the discussed methods to microscopy, which is one of the most intriguing fields of application of plenoptic imaging.

\section{Acknowledgments}

MD, AG and OV acknowledge financial support from the Italian Ministry of Education, University and Research (MIUR), project PON02-00576-3333585 (P.O.N. RICERCA E COMPETITIVIT\`A 2007-2013 - Avviso n. 713/Ric. del 29/10/2010, Titolo II - ``Sviluppo/Potenziamento di DAT e di LPP''). MD, AG and FVP are partially supported by Istituto Nazionale di Fisica Nucleare (INFN) through the projects ``QUANTUM'' and ``PICS''.

\end{document}